\documentclass[a4paper,12pt]{article}
\usepackage{amsmath}
\usepackage{accents}
\usepackage{bm}
\usepackage[english]{babel}
\usepackage{graphicx}
\usepackage[usenames]{color}
\usepackage{colortbl}
\usepackage[toc,page]{appendix}
\usepackage{subcaption}
\usepackage{floatrow}
\begin{document}

\centerline {\LARGE{Geometry and speed of evolution}}
\centerline {\LARGE{for a spin-$s$ system}}
\centerline {\LARGE{with long-range $zz$-type Ising interaction}}
\medskip
\centerline {Yu. S. Krynytskyi$^1$, A. R. Kuzmak$^2$}
\centerline {\small \it E-Mail: yurikryn@gmail.com$^1$, andrijkuzmak@gmail.com$^2$}
\medskip
\centerline {\small \it Department for Theoretical Physics, Ivan Franko National University of Lviv,}
\medskip
\centerline {\small \it 12 Drahomanov St., Lviv, UA-79005, Ukraine}
\medskip

{\small

We study the evolution of a spin-$s$ system described by the long-range $zz$-type Ising interaction.
The Fubini-Study metric of the quantum state manifold defined by this evolution is obtained. We explore the topology of this manifold and show that it
corresponds to a sphere. Exploration of the Riemannian curvature allows us to determine the manifold geometry. Also we calculate the speed of evolution
of the system and represent the curvature by means of this speed. This is important for an experimental measurement of the curvature.
The conditions for achieving the minimal and maximal values of the speed of evolution are obtained. Also we examine the geometry of state manifold
and speed of evolution of spin system in the thermodynamic limit. We propose the physical system of methane molecule for application of our considerations.
Finally, we study the influence of an external magnetic field on the metric of state manifold and on the speed of evolution. In this case we obtain
the conditions for achieving the minimal possible speed of evolution. For some predefined initial states the orientations of magnetic fields
to reach the minimal and maximal values of the speed are found.

\medskip

}

\section{Introduction\label{sec1}}

Information about the geometry of a quantum state manifold plays a crucial role in studying properties of quantum systems
\cite{gqev1,AGATQM,gqev2,FSM,gqev3,FSM2,GNARQCS,IGQPTDM,GIGCQPT,IGRG,UWFDSACD,brody2001}. This information often allows to simplify the study of many problems concerning
quantum dynamics \cite{GANQE,RBPDQE,OCGQC,GAQCLB,QCAG,QGDM,GQCQ,KhGlBr,brach,brachass,brach1,OHfST,TMTSPMF,TOSTSS,QBS1,ZNP1,ZNP2,ZNP3,ZNP4,ZNP5},
quantum speed limits \cite{QSLHUPOQC,MSDE,GCINQP,MSDDL,GLBQCM,GGQSL,QSGMBSD},
quantum entanglement \cite{LSPPTQS,GES,GESBSHF,OGES,GEMCGP,GPEBI,BGESTQ,GSESSV,Entandgeom,torus,FMM,EQSGSSARII},
quantum computations \cite{OCGQC,GAQCLB,QCAG,QGDM,GQCQ,zanardi1999,pachos2000,falci2000,duan2001,solinas2003,sjoqvist2012,zu2014,qcomp1,qcomp2,qcomp3,qcomp4,qcomp5}, etc.
In \cite{gqev1} it was shown that the distance which the system
passes during the quantum evolution along a given curve in the projective Hilbert space is related to the integral of the uncertainty of energy
that in turn is connected with the speed of evolution. Indeed, this distance can be obtained from the definition of the Fubini-Study metric
\cite{AGATQM,gqev2,FSM,FSM2,FSM3,FSM0,FSM1,FSMqsgm,FSMqsgm2}. The Fubini-Study metric is defined by the infinitesimal
distance $ds$ between two neighboring pure quantum states
$\vert\psi (\xi^{\mu})\rangle$ and $\vert\psi (\xi^{\mu}+d\xi^{\mu})\rangle$:
\begin{eqnarray}
ds^2=g_{\mu\nu}d\xi^{\mu}d\xi^{\nu},
\label{form7}
\end{eqnarray}
where $\xi^{\mu}$ is a set of real parameters which define the state $\vert\psi(\xi^{\mu})\rangle$. This state satisfies the following normalization condition
$\langle\psi(\xi^{\mu})\vert\psi(\xi^{\mu})\rangle=1$. Each of these parameters is determined by certain ranges of values. So, the state $\vert\psi(\xi^{\mu})\rangle$
is a function of parameter $\xi^{\mu}$ which take the values from a certain ranges. In turn, the number of these parameters and their ranges define the dimension,
size and topological properties of the quantum state manifold. Each point on this manifold corresponds to a particular quantum state with a predefined set of parameters.
For instance, two real parameters are enough to define an arbitrary state of two-level quantum system. For this purpose, such state is often
parameterized by spherical angles $\theta$, $\phi$ which take the values from the ranges $\theta\in[0,\pi]$, $\phi\in[0,2\pi]$, respectively.
The manifold which contains all states of this system is a sphere with unit radius (Bloch sphere) (see \cite{FSM,FSM2,OHfST,TMTSPMF,TOSTSS}).
The components of the metric tensor $g_{\mu\nu}$ which are present in formula (\ref{form7}) have the form
\begin{eqnarray}
g_{\mu\nu}=\gamma^2\Re\left(\langle\psi_{\mu}\vert\psi_{\nu}\rangle-\langle\psi_{\mu}\vert\psi\rangle\langle\psi\vert\psi_{\nu}\rangle\right),
\label{form8}
\end{eqnarray}
where $\gamma$ is a scale factor, which is often chosen to have the value of $1$, $\sqrt{2}$ or $2$, and
\begin{eqnarray}
\vert\psi_{\mu}\rangle=\frac{\partial}{\partial\xi^{\mu}}\vert\psi\rangle.
\label{form9}
\end{eqnarray}
The metric defined by expression (\ref{form7}) is valid for quantum systems of different dimensionality of Hilbert space including infinite-dimensional
case \cite{AGATQM}. It is important to note that in paper \cite{CMGQGSM} the methods for measuring the metric tensor of a quantum ground state
manifold was proposed. As an example, the results were obtained for the quantum $XY$ chain in a transverse magnetic field. Recently, the generic protocol
to experimentally measure the quantum metric tensor was proposed in paper \cite{EQMTTPD}.

So, the distance which the system passes during the period $t$ is defined by equation
\begin{eqnarray}
s=\int_{0}^{t}\sqrt{g_{t't'}}dt',
\label{pass}
\end{eqnarray}
where
\begin{eqnarray}
g_{t't'}=\gamma^2\langle\psi(t')\vert\left(\Delta H\right)^2\vert\psi(t')\rangle.
\label{metric}
\end{eqnarray}
Here $\Delta H=H-\langle\psi(t')\vert H\vert\psi(t')\rangle$ is the energy uncertainty and $\vert\psi(t')\rangle$ is the state which the system reaches
during the time $t'$. We set $\hbar=1$, which means that the energy is measured in the frequency units. Finally, the speed of evolution is given
by the Anandan-Aharonov relation \cite{gqev1}
\begin{eqnarray}
v=\sqrt{g_{t't'}}.
\label{speed}
\end{eqnarray}
So, the information about the metric of quantum state manifold which contains all states the system can reach during the evolution allows
to calculate the speed of this evolution. In other words, this speed determines the distance that the system passes between two neighboring quantum states
during a certain period of time. As we can see from equations (\ref{speed}) and (\ref{metric}), it is defined by the Hamiltonian parameters. This fact
allows to measure it on experiment, which, in turn, is important for finding the time of evolution between two predefined quantum states.
As a results, we can estimate the time that the system spends to reach the required state.
Also, it is important to note that the states, which the system can reach during the evolution, are defined by the initial state and the form of the Hamiltonian.
These conditions set the quantum state manifold of this system. Thus, for the preparation of a required quantum state on a particular system,
it is important to find the speed of evolution and to set the structure of the state manifold. This knowledge is useful in the study of different
problems of quantum mechanics we mention at the beginning of this section.

In our previous paper \cite{EQSGSSARII} we considered the spin-$1/2$ system described by the long-range $zz$-type Ising model. The topology and geometry
of the quantum state manifold of this system was studied. Also we investigated the entanglement of the states, which belong to this manifold, and
obtained relation between the scalar curvature of the manifold and value of entanglement. In the present paper we consider the spin-$s$ system described
by the $zz$-type Ising Hamiltonian with long-range interaction (Section \ref{sec2}). We study the geometry of the manifold which contains all states the system
can reach during the evolution having started from the initial state projected along the positive direction of some unit vector (Section \ref{sec3}).
We show that this manifold has the topology of sphere. In Section \ref{sec4} the speed of evolution is examined. As a result, the curvature
of the manifold is expressed by the speed of evolution. Also we examine the geometry of the quantum state manifold and the speed of evolution of the spin system
in the thermodynamic limit (Section \ref{sec5}). The results are considered on the physical system of the methane molecule (Section \ref{sec6}).
So, for this system we obtain the behavior of the speed of evolution and dependence of the scalar curvature of the state manifold on this speed.
Finally, we examine the influence of the external magnetic field on metric of the manifold (Section \ref{sec7}).
Also we obtain the conditions for achieving the minimal possible speed of evolution for a predefined initial state. The orientations of the magnetic field
for reaching the minimal and maximal values of speed for some initial states are found. Conclusions are presented in Section \ref{sec8}.

\section{The long-range Ising-type model \label{sec2}}

We consider the $N$ spin-$s$ system with long-range interaction described by the $zz$-type Ising Hamiltonian
\begin{eqnarray}
H=2J\sum_{1\leq i < j\leq N}S_i^zS_j^z,
\label{form1}
\end{eqnarray}
where $J$ is the interaction coupling, $N$ is the number of spins, and $S_i^z$ is the $z$-component of the spin operator which is defined by the following
equation
\begin{eqnarray}
S_i^z\vert m_i\rangle=m_i\vert m_i\rangle.
\label{form2}
\end{eqnarray}
Here $m_i=-s,-s+1,\ldots,s$ are the values of projection of the $i$th spin on $z$ axis, and $\vert m_i\rangle$ are the corresponding eigenstates.
Also it should be noted that the components of spin-$s$ operators $S_i^x$, $S_i^y$ and $S_i^z$ satisfy the following algebra
\begin{eqnarray}
\left[S_i^{\alpha},S_j^{\beta}\right]=i\delta_{ij}\sum_{\gamma=x,y,z}\epsilon^{\alpha\beta\gamma}S_i^{\gamma},
\label{form2_1}
\end{eqnarray}
where $\epsilon^{\alpha\beta\gamma}$ denotes the Levi-Civita symbol and $\delta_{ij}$ is the Kronecker delta. The spin components $S_i^x$ and $S_i^y$
act on eigenstate $\vert m_i\rangle$ of $S_i^z$ as follows
\begin{eqnarray}
&&S_i^x\vert m_i\rangle=\frac{1}{2}\left(\sqrt{s(s+1)-m_i(m_i+1)}\vert m_i+1\rangle \right.\nonumber\\
&&\left.+\sqrt{s(s+1)-m_i(m_i-1)}\vert m_i-1\rangle\right),\nonumber\\
&&S_i^y\vert m_i\rangle=\frac{1}{2i}\left(\sqrt{s(s+1)-m_i(m_i+1)}\vert m_i+1\rangle \right.\nonumber\\
&& \left.-\sqrt{s(s+1)-m_i(m_i-1)}\vert m_i-1\rangle\right),
\label{spincomp}
\end{eqnarray}
where we use the representation of spin component operators through the ladder operators as follows $S_i^x=1/2\left(S_i^++S_i^-\right)$,
$S_i^y=1/(2i)\left(S_i^+-S_i^-\right)$. The eigenstates of Hamiltonian (\ref{form1}) are in general superpositions of all states $\vert m_i\rangle$.

The arbitrary state of the $N$ spin-$s$ system can be decomposed by the eigenstates which satisfy equation (\ref{form2}) as follows
\begin{eqnarray}
\vert\psi\rangle=\sum_{m_1,m_2,\ldots,m_N}a_{m_1,m_2,\ldots,m_N}\vert m_1,m_2,\ldots,m_N\rangle
\label{form3}
\end{eqnarray}
with the following normalization condition $\sum_{m_1,m_2,\ldots,m_N}\vert a_{m_1,m_2,\ldots,m_N}\vert^2=1$.
Then the evolution of this state under the action of Hamiltonian (\ref{form1}) takes the form
\begin{eqnarray}
e^{-iHt}\vert\psi\rangle=\sum_{m_1,m_2,\ldots,m_N}a_{m_1,m_2,\ldots,m_N}e^{-i2\chi\sum_{1\leq i < j\leq N}m_im_j}\vert m_1,m_2,\ldots,m_N\rangle.
\label{form4}
\end{eqnarray}
Here $\chi=Jt$ is a variable parameter which depends on $t$, because interaction coupling $J$ has a predefined value.
From the analysis of state (\ref{form4}) we obtain that it is periodic with respect to parameter $\chi\in[0,\chi_{max}]$, where
$\chi_{max}=2\pi$ for half-integer $s$ and $\chi_{max}=\pi$ for integer $s$. So, let us study the geometry of a manifold which contains the states reached
during the evolution of system having started from the initial state, where all spins have maximal values $s$ of projection on the positive direction of the unit vector
${\bf n}=\left(\sin\theta\cos\phi,\sin\theta\sin\phi,\cos\theta\right)$. This means that all spins of the system in the initial state are directed along the positive direction
of unit vector ${\bf n}$ and have a maximal value of projection. Here $\theta$ and $\phi$ are the polar and azimuthal angles, respectively.
This state is called a polarized product state. To simplify further calculations we express this state as follows
\begin{eqnarray}
\vert\psi_I\rangle=e^{-i\phi\sum_{i=1}^NS_i^z}e^{-i\theta\sum_{i=1}^NS_i^y}\vert s,s,\ldots,s\rangle.
\label{form5}
\end{eqnarray}
It can be easily prepared when the spin system is placed in the strong magnetic field ($h\gg J$) directed along  the unit vector ${\bf n}$,
where $h$ is the value of interaction between the magnetic field and each spin (see, for instance, \cite{phosphorus1}). The ground state of such system
has the required form (\ref{form5}). Then the evolution of the system can be expressed as follows
\begin{eqnarray}
\vert\psi(t)\rangle=e^{-i2\chi\sum_{1\leq i < j\leq N}S_i^zS_j^z}e^{-i\phi\sum_{i=1}^NS_i^z}e^{-i\theta\sum_{i=1}^NS_i^y}\vert s,s,\ldots,s\rangle.
\label{form6}
\end{eqnarray}

\section{The geometry of quantum state manifold \label{sec3}}

\begin{figure}[!!h]
\centerline{\includegraphics[scale=0.5, angle=0.0, clip]{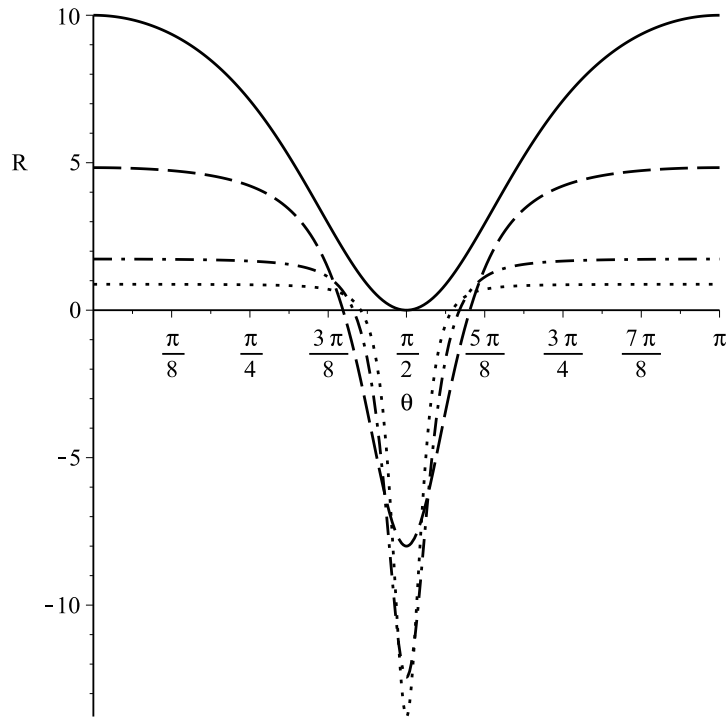}}
\caption{{\it The dependence of scalar curvature on $\theta$ (\ref{curvature2}). Results are presented for different
numbers and values of spins:  $N=2$, $s=1/2$ (solid curve), $N=3$, $s=1$ (dashed curve), $N=6$, $s=3/2$ (dash-dotted curve) and $N=9$, $s=2$ (dotted curve).
Here $\gamma=1$.}}
\label{curvature_as}
\end{figure}

To study the geometry of quantum state manifold we calculate its Fubini-Study metric. For this purpose we use definition (\ref{form8}) for state
(\ref{form6}) with respect to parameters $\chi$, $\theta$ and $\phi$. So, using the fact
\begin{eqnarray}
e^{i\theta\sum_{j=1}^NS_j^y}S_i^ze^{-i\theta\sum_{j=1}^NS_j^y}=S_i^z\cos\theta-S_i^x\sin\theta,
\label{BCHeq}
\end{eqnarray}
which follows from the Baker-Campbell-Hausdorff formula and properties of spin operator algebra (\ref{form2_1}) we calculate the components of metric
tensor
\begin{eqnarray}
&&g_{\theta\theta}=\gamma^2\frac{Ns}{2},\quad g_{\phi\phi}=\gamma^2\frac{Ns}{2}\sin^2\theta,\nonumber\\
&&g_{\chi\chi}=\gamma^2N(N-1)s^2\sin^2\theta\left[2s\left(N-1\right)-\left(2s\left(N-1\right)-\frac{1}{2}\right)\sin^2\theta\right],\nonumber\\
&&g_{\theta\phi}=0,\quad g_{\theta\chi}=0,\quad g_{\phi\chi}=\gamma^2N(N-1)s^2\cos\theta\sin^2\theta.
\label{form10}
\end{eqnarray}
The detailed derivation of this metric is presented in Appendix \ref{appa}.
As we can see, the components of metric tensor depend only on parameter $\theta$. Let us explore the geometry of this manifold in detail.
If the parameter $\chi$ is fixed then we have the manifold which contains all possible initials states. It has the geometry of a sphere with radius
$\gamma\sqrt{Ns/2}$ (for instance, see, \cite{brachass,coherentstate1,coherentstate2}).
Also it is easy to see that the manifolds with different certain $\phi$ have the same geometry.
These manifolds are defined by two parameters $\theta$ and $\chi$. They contain all the states which the system reaches during the evolution.
To analyse the topology of these manifolds we calculate their scalar curvature. Using the fact that they have a two-parametric diagonal metric,
which depends only on parameter $\theta$, we represent the scalar curvature in the form \cite{CTF}
\begin{eqnarray}
R=\frac{2R_{\theta\chi\theta\chi}}{g_{\theta\theta}g_{\chi\chi}},
\label{curvature1}
\end{eqnarray}
where
\begin{eqnarray}
R_{\theta\chi\theta\chi}=-\frac{1}{2}\frac{\partial^2 g_{\chi\chi}}{\partial\theta^2}+\frac{1}{4g_{\chi\chi}}\left(\frac{\partial g_{\chi\chi}}{\partial\theta}\right)^2
\label{riemanntensor}
\end{eqnarray}
is the Riemann curvature tensor. Now we substitute the components of metric tensor (\ref{form10}) into (\ref{curvature1}) and after
simplifications obtain that
\begin{eqnarray}
R=\frac{8}{\gamma^2Ns}\left(2-\frac{(4(N-1)s-1)\cos^2\theta+2(N-1)s+1}{\left[(4(N-1)s-1)\cos^2\theta+1\right]^2}\right).
\label{curvature2}
\end{eqnarray}
As we can see from this expression, for $N>2$ and $s=1/2$ or for $s>1/2$ the areas with negative curvature with minimum
at $\theta=\pi/2$ appears on the manifold. The minimal value of curvature is determined by the expression
\begin{eqnarray}
R_{min}=\frac{8}{\gamma^2Ns}(1-2(N-1)s).
\label{curvaturemin}
\end{eqnarray}
Near the points with $\theta=0$ and $\pi$ we obtain the maximal value of the curvature. For the case of $N>2$ and $s>1/2$ we do not know anything about
the values of curvature in these point because they are singular and the manifold at them is not a differentiable. However, in all other points it is differentiable.
Also the dependence of $R$ on $\theta$ is symmetric with respect to minimum of the curvature. For example, the dependence of the scalar curvature
for some $N$ and $s$ is shown in Figure \ref{curvature_as}.

Now let us determine the topology of state manifold. For this purpose we use the Gauss-Bonnet theorem and taking into account that for $N>2$ and $s>1/2$
this manifold has an angular defects near the points with $\theta=0$ and $\pi$. In this case the Gauss-Bonnet theorem takes the form
\begin{eqnarray}
\int_{M}\frac{R}{2}\sqrt{g}dA+\Delta=2\pi X(M),
\label{gaussbonnet}
\end{eqnarray}
where $\sqrt{g}dA$ is the element of area of the manifold $M$, $g$ is the determinant of the metric tensor, $\Delta$ defines the contribution of angular
defects and $X(M)$ is the Euler characteristic of the manifold. Using the fact that $\theta\in[0,\pi]$, $\chi\in[0,\chi_{max}]$ the integral in formula (\ref{gaussbonnet})
takes the value $4\chi_{max}(N-1)s$. The angular defects are located very close to the point $\theta=0$ and $\pi$. This fact allows
to rewrite the metric in these areas as follows
\begin{eqnarray}
g_{\theta\theta}=\gamma^2\frac{Ns}{2},\quad g_{\chi\chi}=2\gamma^2N(N-1)^2s^3\theta^2,\quad g_{\theta\chi}=0
\label{form11}
\end{eqnarray}
and obtain that
\begin{eqnarray}
\Delta=2\left(2\pi-\frac{\sqrt{g_{\chi\chi}}\chi_{max}}{\sqrt{g_{\theta\theta}}\theta}\right)=2\left(2\pi -2(N-1)s\chi_{max}\right).
\label{form12}
\end{eqnarray}
Multiplier 2 means that the manifold has two angular defects. The detailed derivation of formula (\ref{form12}) is presented in Appendix \ref{appb}.
The components of the metric tensor which appear in equation (\ref{form12}) are defined by expressions (\ref{form11}). Substituting $\Delta$ into Gauss-Bonnet formula (\ref{gaussbonnet})
we obtain the Euler characteristic $X(M)=2$. This means that the manifold has the topology of the sphere. Taking into account all the above
we conclude that the manifold described by metric (\ref{form10}) with respect to parameters $\theta\in[0,\pi]$ and $\chi\in[0,\chi_{max}]$ is closed
and has a dumbbell-shape structure for $N>2$ and $s=1/2$ or for $s>1/2$. It has the concave part with the centerline at $\theta=\pi/2$
and is symmetric with respect to it.

\section{Speed of evolution \label{sec4}}

Now using definition of the speed of evolution (\ref{speed}) with $"\chi\chi"$ component of the metric tensor we obtain the equation for the speed
of evolution of the spin system defined by Hamiltonian (\ref{form1}) in the form
\begin{eqnarray}
v=\vert J\vert\sqrt{g_{\chi\chi}}.
\label{form13}
\end{eqnarray}
Note that the speed of evolution can contain terms which are proportional to $N^2$. This fact follows
from definitions (\ref{metric}), (\ref{speed}) and structure of Hamiltonian (\ref{form1}). However, in our case the maximal power
of $N$ is $3/2$. This is because the terms with $N^2$ cancel, which is provided by the structure of scalar products $\langle\psi\vert\psi_{\chi}\rangle$
and $\langle\psi_{\chi}\vert\psi_{\chi}\rangle$ (see expression (\ref{scalarprod}) in Appendix \ref{appa}).
We can see that this speed depends on parameter $\theta$ and does not depend on parameters $\chi$ and $\phi$.
The distance which the system passes during the evolution is the following
\begin{eqnarray}
s=\sqrt{g_{\chi\chi}}\chi.
\label{form14}
\end{eqnarray}

\begin{figure}[!!h]
\centerline{\includegraphics[scale=0.5, angle=0.0, clip]{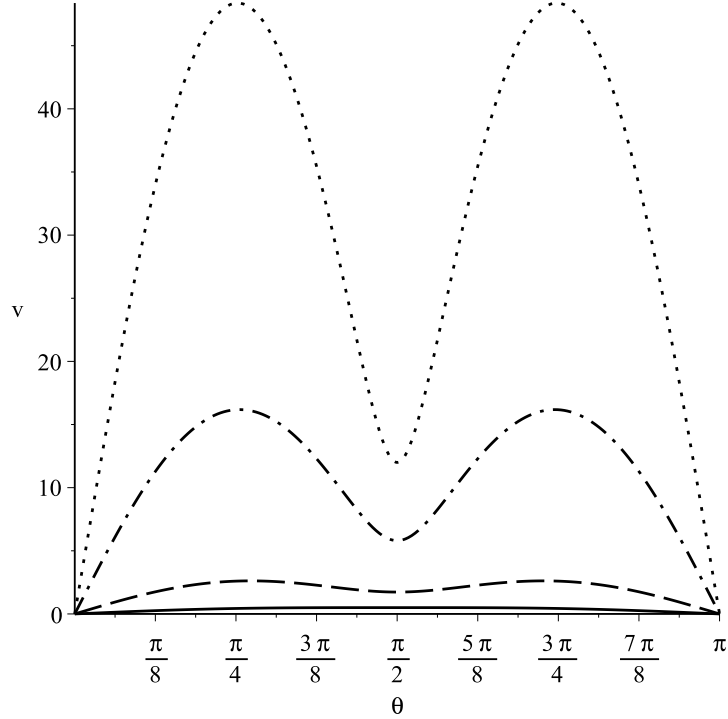}}
\caption{{\it The dependence of the speed of evolution on $\theta$ (\ref{form13}). Results are presented for different
numbers and values of spins:  $N=2$, $s=1/2$ (solid curve), $N=3$, $s=1$ (dashed curve), $N=6$, $s=3/2$ (dash-dotted curve) and $N=9$, $s=2$ (dotted curve).
Here $\gamma=1$ and $J=1$.}}
\label{speed_as}
\end{figure}

\begin{figure}[!!h]
\centerline{\includegraphics[scale=0.5, angle=0.0, clip]{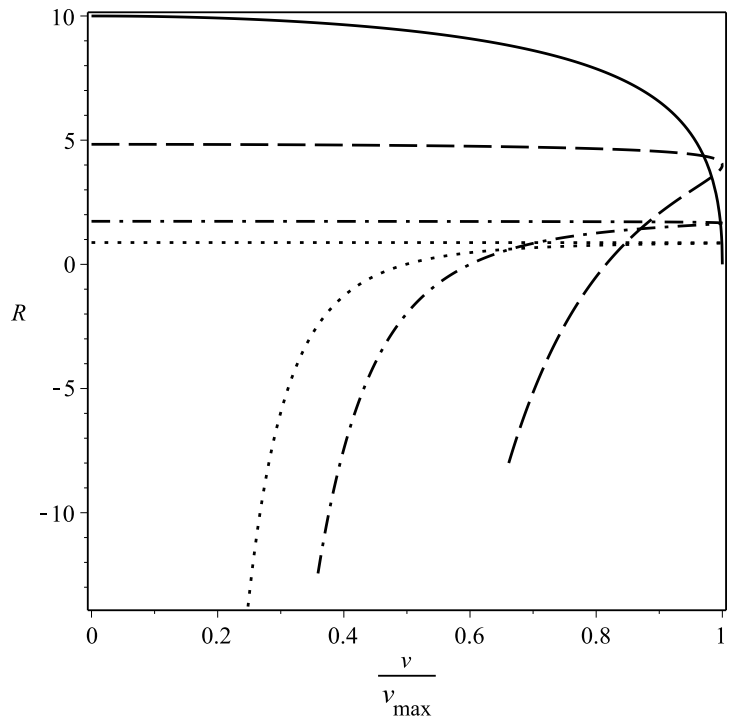}}
\caption{{\it The dependence of the scalar curvature on speed of evolution. Results are presented for different
numbers and values of spins:  $N=2$, $s=1/2$ (solid curve), $N=3$, $s=1$ (dashed curve), $N=6$, $s=3/2$ (dash-dotted curve) and $N=9$, $s=2$ (dotted curve).
Here $\gamma=1$.}}
\label{curvature_speed_as}
\end{figure}

Let us calculate the maximal and minimal values of the speed. For this purpose we find the extremum of $g_{\chi\chi}$ with respect to $\theta$.
Using the fact that $\theta\in\left[0,\pi\right]$ we obtain that at the points $\theta=0$ and $\pi$ the speed has
the minimal value $v_{min}=0$, at the point $\theta=\pi/2$ it has the local minimum
\begin{eqnarray}
v_{\pi/2}=\vert J\vert\gamma s\sqrt{\frac{N(N-1)}{2}}
\label{form15_1}
\end{eqnarray}
and at the two symmetric points which are defined by the equation
\begin{eqnarray}
\sin^2\theta_{max}=\frac{(N-1)s}{2(N-1)s-1/2}
\label{form16}
\end{eqnarray}
it has the maximum
\begin{eqnarray}
v_{max}=\vert J\vert\gamma (N-1)s^2\sqrt{\frac{N(N-1)}{2(N-1)s-1/2}}.
\label{form17}
\end{eqnarray}
However if $N=2$ and $s=1/2$ the speed has one maximum at $\theta=\pi/2$. The dependence of the speed of evolution on parameter $\theta$
for different numbers and values of spins is shown in Figure \ref{speed_as}.

Using the fact that the speed of evolution (\ref{form13}) and Riemannian curvature (\ref{curvature2}) depend on the parameter $\theta$ we can express
the curvature by the speed $v$. In turn this means that the curvature of manifold can be experimentally measured because the speed of evolution
is defined by the energy uncertainty (see (\ref{speed}) with (\ref{metric})). So, from equation (\ref{form13}) with (\ref{form10}) we obtain
\begin{eqnarray}
\sin^2\theta=\sin^2\theta_{max}\left(1\mp\sqrt{1-\left(\frac{v}{v_{max}}\right)^2}\right).
\label{form18}
\end{eqnarray}
The expression with upper sign corresponds to the ranges $\theta\in[0,\theta_{max}]$, $\theta\in[\pi-\theta_{max},\pi]$, when the speed of evolution
varies as follows $v\in[0,v_{max}]$, and the expression with lower sign corresponds to the range $\theta\in[\theta_{max},\pi-\theta_{max}]$,
when the speed of evolution varies as follows $v\in[v_{max},v_{\pi/2}]$, $v\in[v_{\pi/2},v_{max}]$.
Then substituting these expressions into (\ref{curvature2}) we obtain the dependence of the scalar curvature on the speed of evolution for the first and second ranges
\begin{eqnarray}
R=\frac{8}{\gamma^2Ns}\left[2-\frac{2\pm\sqrt{1-\left(\frac{v}{v_{max}}\right)^2}}{2(N-1)s\left(1\pm\sqrt{1-\left(\frac{v}{v_{max}}\right)^2}\right)^2}\right],
\label{form20}
\end{eqnarray}
respectively. The dependence of the scalar curvature on the speed of evolution for some $N$ and $s$ is depicted in Figure \ref{curvature_speed_as}.

\section{Geometry and speed of evolution of the spin system in the thermodynamic limit \label{sec5}}

It is important to consider the behavior of a many-body system in thermodynamic limit. Therefore, in this section we study the geometry of
the quantum state manifold and speed of evolution of the system defined by Hamiltonian (\ref{form1}) in the case of $N\to\infty$. For this purpose we divide the coupling constant $J$
in Hamiltonian (\ref{form1}) by the number of spins $N$, so that the energy per spin of the system remains finite in thermodynamic limit.
Then the metric of the system is defined by expression (\ref{form10}) with new $g_{\chi\chi}$ component, which additionally has a multiplier $1/N^2$,
and $g_{\phi\chi}$ component, which additionally has a multiplier $1/N$. The rescaling of the coupling
constant does not influence the scalar curvature of the manifold. Thus, expression (\ref{curvature2}) also describes the Riemannian curvature in the
thermodynamic limit. It is easy to see that if $N\to\infty$ then the scalar curvature tends to zero ($R\to 0$) at all points of the state manifold
except the points with $\theta=\pi/2$. This is due to the fact that an increase in the number of spins in the system leads to the inflation
of the quantum state manifold and it becomes flattened.
Also the dumbbell-shape structure of the manifold provides the concave part with the centerline at $\theta=\pi/2$ which in the thermodynamic limits
narrows to the closed line at $\theta=\pi/2$. Then the curvature on this line takes the value $-16/\gamma^2$. In fact, this line is the boundary
between two very inflated parts of a state manifold. Because of the value of spin $s$ is presented in the numerator and denominator of the second
term of expression (\ref{curvaturemin}) the curvature in the thermodynamic limit does not depend on $s$. Also it is worth noting that similar
results we obtain in the case when the value of spin tends to infinity ($s\to\infty$). However, in this case the limit value of curvature on
the line with $\theta=\pi/2$ depends on the number of spins as follows: $R=-16(N-1)/(\gamma^2 N)$. Then, depending on the number of spins in system,
it takes the values $R\in[-8/\gamma^2,-16/\gamma^2)$. We assume that the minimal number of spins in the system can be taken as $N_{min}=2$.

The speed of evolution is defined by equation (\ref{form13}) with additionally multiplier $1/N$. Then, in the thermodynamic limit it diverges
as\linebreak $v=\vert J\vert\gamma s^{3/2}\sqrt{N}\sin(2\theta)/\sqrt{2}$ for all $\theta$ except $\theta=\pi/2$. At $\theta=\pi/2$, the speed of evolution tends
to $v_{\pi/2}=\vert J\vert\gamma s/\sqrt{2}$. Also, the maximal value of the speed in the thermodynamic limit has the form
$v_{max}=\vert J\vert\gamma s^{3/2}\sqrt{N}/\sqrt{2}$. The points on the manifold corresponding to this value are given by $\theta_{max}=\pi/4$ and $3\pi/4$.

\section{Application to the physical system of the methane molecule \label{sec6}}

The results which we obtain in the present paper can be used for the exploration of quantum state manifolds of different physical systems.
Information about the geometry properties of this manifolds are essential for implementation of quantum computations
\cite{OCGQC,GAQCLB,QCAG,QGDM,GQCQ,zanardi1999,pachos2000,falci2000,duan2001,solinas2003,sjoqvist2012,zu2014}.
The authors of papers \cite{OCGQC,GAQCLB,QCAG,QGDM,GQCQ} showed that the problem of finding an optimal quantum circuit of a unitary operation is closely related to the problem
of finding the minimal distance between two points on the Riemannian metric. Using the geometric properties of quantum systems
the experimental implementation of quantum gates was proposed in papers \cite{falci2000,duan2001,solinas2003}. In paper \cite{zu2014} the authors
report the experimental realization of universal geometric quantum gates using the solid-state spins of diamond nitrogen-vacancy centers.
We find the connection between the speed of evolution and the scalar curvature of the spin-$s$ $zz$-type Ising system with long range interaction.
So, measurement of the speed of evolution of the spin system provides the information about the curvature of its quantum state manifold. Let us apply the above results to real physical system.
\begin{figure}[!!h]
\centerline{\includegraphics[scale=1, angle=0.0, clip]{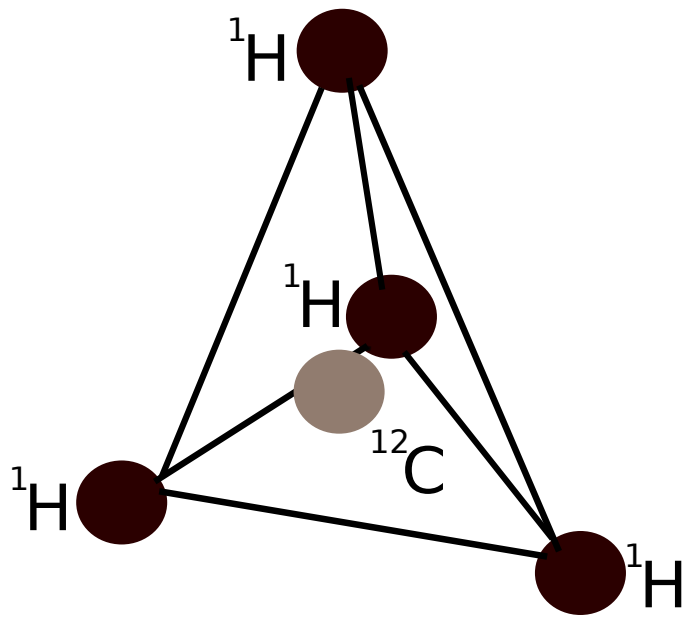}}
\caption{{\it Structure of the methane molecule. The nuclear spins of the hydrogen atoms interact between themselves and do not interact with carbon nucleus.}}
\label{fivespincluster}
\end{figure}
\begin{figure}[!!h]
\subcaptionbox{\label{}}{\includegraphics[scale=0.34, angle=0.0, clip]{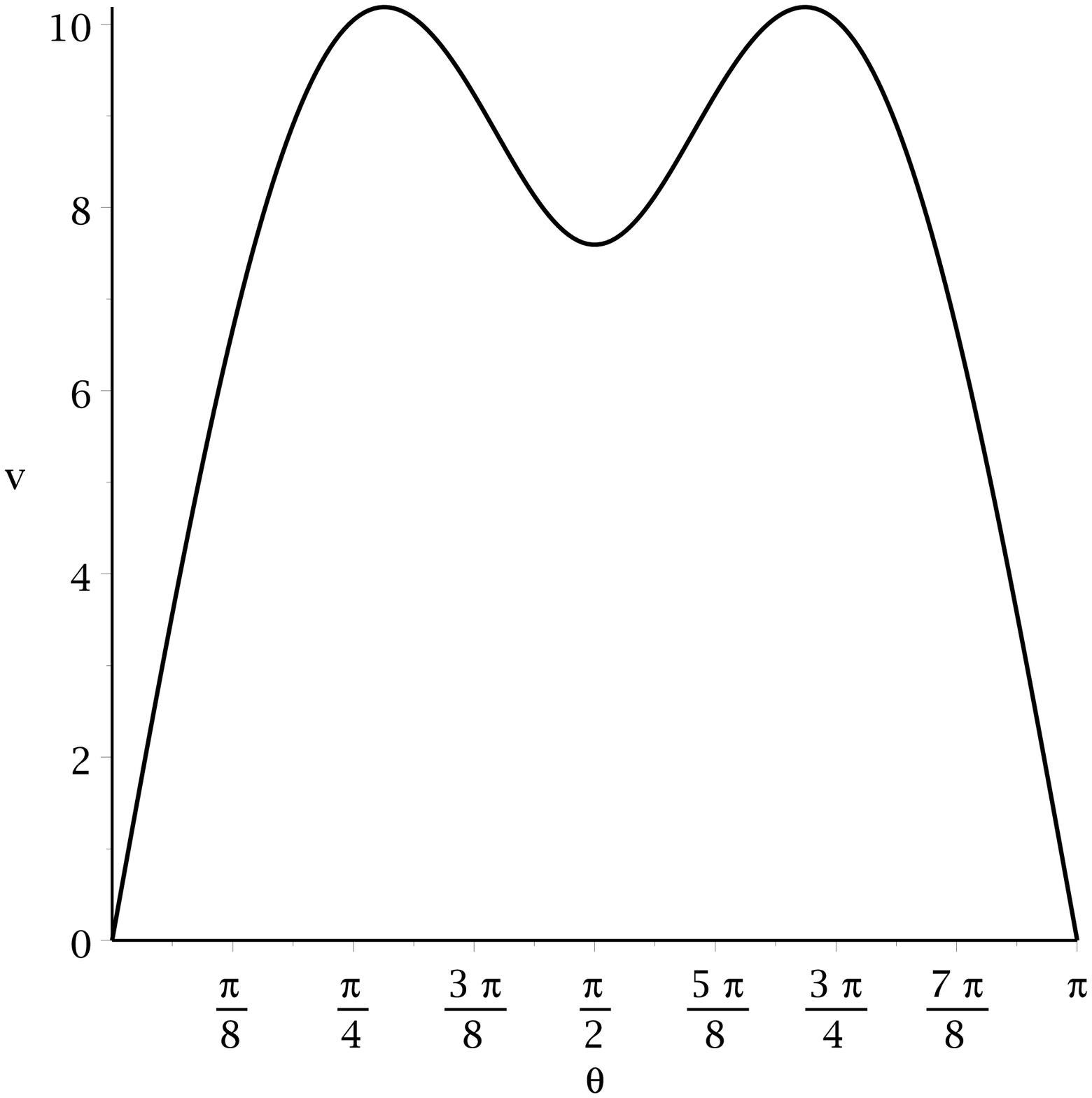}}
\subcaptionbox{\label{}}{\includegraphics[scale=0.34, angle=0.0, clip]{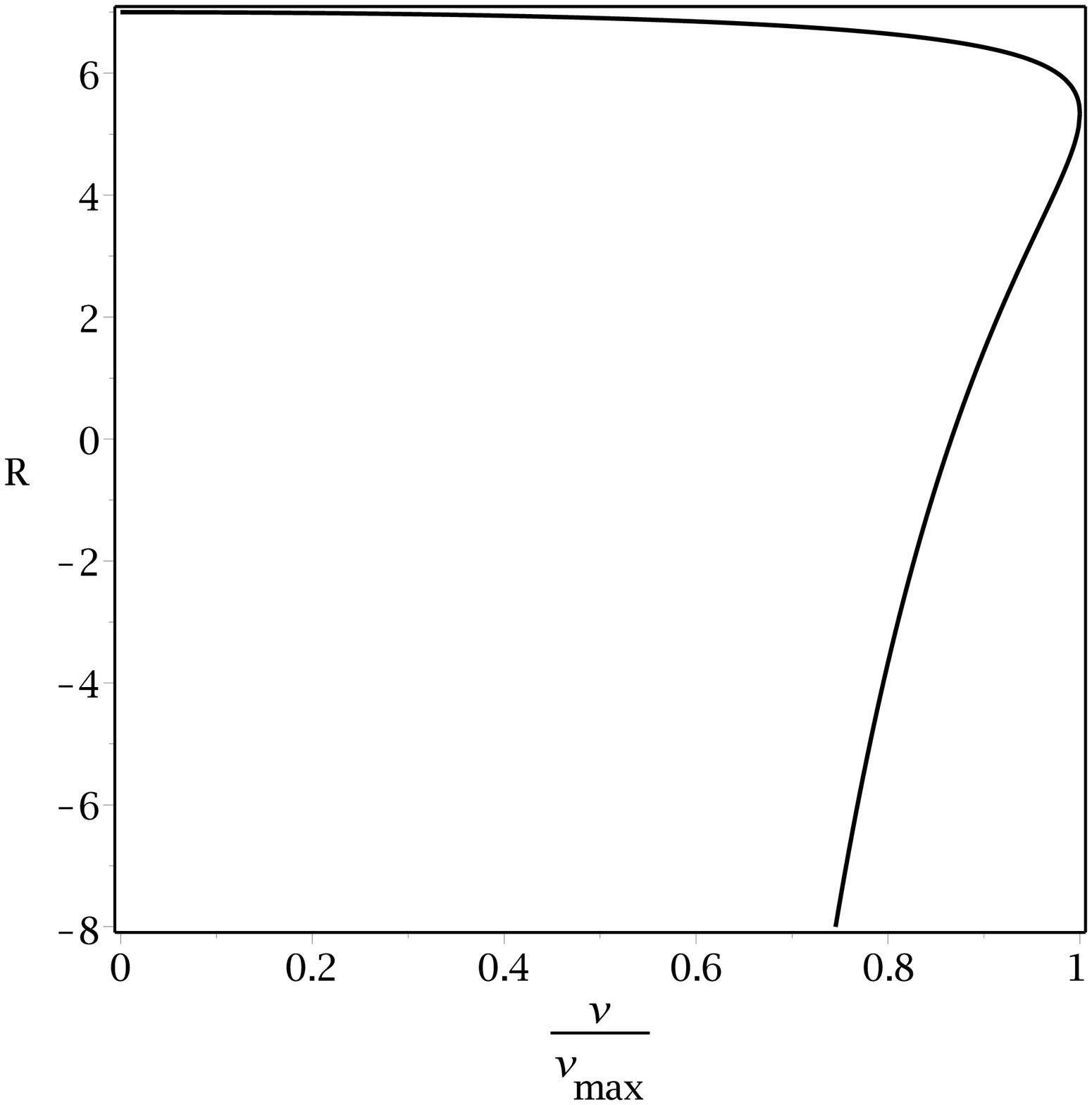}}
\caption{The behavior of the speed of evolution (a) and the dependence of the scalar curvature
of quantum state manifold on speed of evolution (b) of four spin-$1/2$ system in the methane molecule.}
\label{methanesc}
\end{figure}

We suggest the system of the nuclear spins of hydrogen atoms in the methane molecule for experimental realization of our considerations.
The methane molecule consists of four atoms of $^1$H and one atom of $^{12}$C (Figure \ref{fivespincluster}.). Each nucleus of the hydrogen atom consists
of one proton and has spin $1/2$. The nucleus of $^{12}$C isotope has spin 0. The structure of the methane molecule is tetrahedral
(see Figure \ref{fivespincluster}). Therefore, the nuclear spins of hydrogen atoms mutually interact.
We have the system of four spin-$1/2$, where all spins interact between themselves. We assume that this interaction is described by Ising model
(\ref{form1}) with $J=-6.2$ Hz. The value of interaction coupling between proton spins in methane was found in paper \cite{NMRCC}.
So, having started from initial state (\ref{form5}) all the states reached during the evolution of this system are located on the manifold
with metric (\ref{form10}), where $N=4$ and $s=1/2$. The scalar curvature of two-parametric submanifold defined by parameters $\chi$ and $\theta$
is the following
\begin{eqnarray}
R=\frac{4}{\gamma^2}\left(2-\frac{5\cos^2\theta+4}{\left(5\cos^2\theta+1\right)^2}\right)
\label{methanecurvature}
\end{eqnarray}
with minimum value $-8/\gamma^2$ at $\theta=\pi/2$. The relation between the curvature and speed of evolution has form (\ref{form20}) with $N=4$, $s=1/2$
and $v_{max}\approx 10.19\gamma$ Hz. In Figure \ref{methanesc} we show the behavior of the speed of evolution (a) and the dependence of the manifold curvature
on this speed (b). It is important to note that similar properties have the following systems: SiH$_4$, GeH$_4$ and SnH$_4$.

\section{The long-range Ising model in the arbitrary magnetic field \label{sec7}}

In this section we study the influence of the arbitrary magnetic field on the metric of quantum state manifold and
on the speed of evolution of the spin system. The Hamiltonian of such a system has the form
\begin{eqnarray}
H=2J\sum_{1\leq i< j\leq N}S_i^zS_j^z+h\sum_{j=1}^{N}{\bf S}_j \cdot {\bf n'},
\label{form21}
\end{eqnarray}
where $h$ is proportional to the strength of the magnetic field and the unit vector ${\bf n'}=\left(\sin\theta'\cos\phi',\sin\theta'\sin\phi',\cos\theta'\right)$
defines the direction of the magnetic field.

\begin{figure}[!!h]
\centerline{\includegraphics[scale=0.5, angle=0.0, clip]{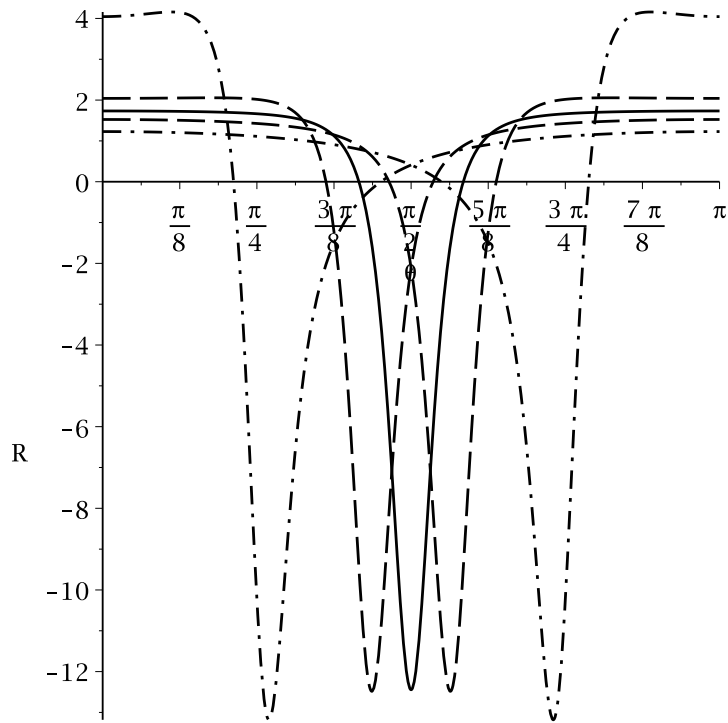}}
\caption{The influence of magnetic field on scalar curvature for the case when magnetic field is directed along the $z$-axis. Results are presented for
six spin-$3/2$ system and different values of magnetic fields: $h/J=0$ (solid curve), $h/J=3$ (dashed curve) and $h/J=10$ (dash-dotted curve).
The magnetic field deforms the manifold. In the case of $J>0$ the concavity is shifted to the right or left side when the magnetic field is directed along the positive
or negative direction of $z$-axis, respectively, and vice versa in the case of $J<0$. Here $\gamma=1$.}
\label{curvature_as_mf}
\end{figure}

So, the evolution of the system having started from state (\ref{form5}) can be expressed in the following form
\begin{eqnarray}
\vert\psi\rangle=\exp{\left\{-i2\chi\left(\sum_{1\leq i<j\leq N}S_i^zS_j^z+\frac{h}{2J}\sum_{j=1}^{N}{\bf S}_j\cdot {\bf n'}\right)\right\}}\vert\psi_I\rangle.
\label{form22}
\end{eqnarray}
In the same way as is described in Appendix \ref{appa} we calculate the Fubini-Study metric in this case. Then we obtain the metric tensor in the form
\begin{eqnarray}
&&g_{\theta\theta}=\gamma^2\frac{Ns}{2},\quad g_{\phi\phi}=\gamma^2\frac{Ns}{2}\sin^2\theta,\nonumber\\
&&g_{\chi\chi}=\gamma^2N(N-1)s^2\sin^2\theta\left[2s\left(N-1\right)-\left(2s\left(N-1\right)-\frac{1}{2}\right)\sin^2\theta\right]\nonumber\\
&&+\gamma^2\left(\frac{h}{J}\right)^2\frac{Ns}{2}\left[\left({\bf n'}\cdot{\bf n}(\theta+\pi/2)\right)^2+\left({\bf n'}\cdot{\bf n}(\theta=\pi/2,\phi+\pi/2)\right)^2\right]\nonumber\\
&&-2\gamma^2\frac{h}{J}N(N-1)s^2{\bf n'}\cdot{\bf n}(\theta+\pi/2)\cos\theta\sin\theta,\nonumber\\
&&g_{\theta\phi}=0,\quad g_{\theta\chi}=\gamma^2\frac{h}{J}\frac{Ns}{2}{\bf n'}\cdot{\bf n}(\theta=\pi/2,\phi+\pi/2),\nonumber\\
&&g_{\phi\chi}=\gamma^2N(N-1)s^2\cos\theta\sin^2\theta-\gamma^2\frac{h}{J}\frac{Ns}{2}{\bf n'}\cdot{\bf n}(\theta+\pi/2)\sin\theta.
\label{form23}
\end{eqnarray}
where
\begin{eqnarray}
&&{\bf n'}\cdot{\bf n}(\theta+\pi/2)=\cos\theta\sin\theta'\cos(\phi-\phi')-\sin\theta\cos\theta',\nonumber\\
&&{\bf n'}\cdot{\bf n}(\theta=\pi/2,\phi+\pi/2)=-\sin\theta'\sin(\phi-\phi').\nonumber
\end{eqnarray}
are the scalar products between the corresponding vectors. Here ${\bf n}(\theta,\phi)$ is the unit vector with the specific conditions on its coordinates.
We can see that the magnetic field adds the dependence of the metric on the parameter $\phi$. If at least one of the scalar products
${\bf n'}\cdot{\bf n}(\theta+\pi/2)\neq 0$ or ${\bf n'}\cdot{\bf n}(\theta=\pi/2,\phi+\pi/2)\neq 0$ (except the case of $\theta'=0$ or $\pi$) then
we do not know anything about the topology of this manifold because under the influence of the external magnetic field the evolution of the system becomes
very complicated. We can argue that in these cases there are no points on manifold where $g_{\chi\chi}=0$. So, this means that the eigenstates
of the system defined by Hamiltonian (\ref{form21}) do not belong to the manifold. However, we can analyse the case when the magnetic field is directed along the
$z$-axis ($\theta'=0$ or $\pi$). Then ${\bf n'}\cdot{\bf n}(\theta+\pi/2)=\mp\sin\theta$ and ${\bf n'}\cdot{\bf n}(\theta=\pi/2,\phi+\pi/2)=0$,
where upper and lower signs correspond to the cases of $\theta'=0$ and $\pi$, respectively. In this case metric does not depend on parameter $\phi$ which
means that the manifolds with different $\phi$ have the same geometry. Here the periodicity of evolution depends on the ratio $h/J$.
If this ratio is an irrational number the parameter takes values $\chi\in[-\infty,\infty]$ and manifold is open with respect to this parameter.
So, in this case the manifold has the topology of an infinitely long cylinder. In the case if ratio $h/J$ is a rational number $p/q$ then the evolution
is periodic with respect to $\chi\in[0,q\chi_{max}]$, where $p$ and $q$ are coprime integers. Here the manifold has the topology of sphere.
Similarly as in the case considered in the previous sections this manifold also have the angular defects in points with $\theta=0$ and $\pi$. As an example,
for this case the influence of the magnetic field on the scalar curvature of state manifold is shown in Figure \ref{curvature_as_mf}. As we can see
that the magnetic field deforms the manifold. In the case of $J>0$ the concavity is shifted to the right or left side with respect to parameter $\theta$ when
the magnetic field is directed along the positive or negative direction of $z$-axis, respectively, and vice versa in the case of $J<0$.
Also the magnetic field changes the maximal and minimal values of curvature.

Let us study the speed of evolution of the spin  system described by Hamiltonian (\ref{form21}). Similarly as in the previous case the speed and the distance of evolution are determined by expressions (\ref{form13}) and (\ref{form14})
with $g_{\chi\chi}$ from (\ref{form23}), respectively. To obtain the optimal directions of the magnetic field, for achieving
the maximal and minimal values of speed, the component $g_{\chi\chi}$ needs to be explored on an extremum with respect to $\theta'$ and $\phi'$.
From (\ref{form23}) it follows the fact that the higher ratio $h/J$ the greater speed of evolution. Also there is a value of magnetic field which
allows the system with predefined parameters to evolve with the minimal speed. This value is determined by the following expression
\begin{eqnarray}
\left(\frac{h}{J}\right)_{min}=\frac{(N-1)s\sin2\theta{\bf n'}\cdot{\bf n}(\theta+\pi/2)}{\left({\bf n'}\cdot{\bf n}(\theta+\pi/2)\right)^2+\left({\bf n'}\cdot{\bf n}(\theta=\pi/2,\phi+\pi/2)\right)^2}.
\label{form24}
\end{eqnarray}
Then the speed of evolution takes the form
\begin{eqnarray}
&&v_{min}=\vert J\vert\gamma s\sqrt{\frac{N(N-1)}{2}}\left[\sin^4\theta\right.\nonumber\\
&&\left.+(N-1)s\sin^22\theta\frac{\left({\bf n'}\cdot{\bf n}(\theta=\pi/2,\phi+\pi/2)\right)^2}{\left({\bf n'}\cdot{\bf n}(\theta+\pi/2)\right)^2+\left({\bf n'}\cdot{\bf n}(\theta=\pi/2,\phi+\pi/2)\right)^2}\right]^{1/2}.
\label{form25}
\end{eqnarray}
It is easy to see that the minimal possible value of speed we can reach when ${\bf n'}\cdot{\bf n}(\theta=\pi/2,\phi+\pi/2)=0$ that corresponds to
$\theta'=0$, $\pi$ or $\phi'-\phi=0$, $\pi$. So, we obtain that
\begin{eqnarray}
v_{min}=\vert J\vert\gamma s \sqrt{\frac{N(N-1)}{2}}\sin^2\theta
\label{form26}
\end{eqnarray}
and that the value of the magnetic field is
\begin{eqnarray}
\left(\frac{h}{J}\right)_{min}=\frac{(N-1)s\sin2\theta}{{\bf n'}\cdot{\bf n}(\theta+\pi/2)}.
\label{form27}
\end{eqnarray}

Finding the conditions for achieving the maximal value of speed in general case is a very difficult problem. Therefore we consider the some particular cases
of initial states. Let us, as an example, obtain the conditions for achieving the minimal and maximal speeds of evolution for predefined initial states.
First of all we consider the situation when the initial state is projected on the positive $\theta=0$ or negative $\theta=\pi$ direction
of the $z$-axis. Then the speed of evolution has the form $v=\vert J\vert\gamma(h/J)\sqrt{Ns/2}\sin\theta'$. So, we obtain that the minimal speed $v_{min}=0$ is
reached for $\theta'=0$ or $\pi$ and maximal value of speed $v_{max}=\vert J\vert\gamma (h/J)\sqrt{Ns/2}$ is obtained when $\theta'=\pi/2$. In the first case the magnetic
field is directed along the $z$-axis and this means that the initial state is an eigenstate of the system. In the second case the magnetic field located in the orthogonal
plane to the initial states. Another situation we obtain when the initial state is prepared in the $xy$-plane ($\theta=\pi/2$). Then the speed of evolution
takes the form $v=\vert J\vert\gamma\sqrt{Ns/2}\sqrt{(N-1)s+(h/J)^2\left(1-\sin^2\theta'\cos^2(\phi'-\phi)\right)}$. We obtain the minimal speed
$v_{min}=\vert J\vert\gamma s\sqrt{N(N-1)/2}$ if $\theta'=\pi/2$, $\phi'-\phi=0$ or $\pi$, and the maximal speed
$v_{max}=\vert J\vert\gamma\sqrt{Ns/2}\sqrt{(N-1)s+(h/J)^2}$ when $\theta'=0$, $\pi$ or $\phi'-\phi=\pi/2$, $3\pi/2$. Finally, let us consider the situation
when $\theta=\pi/4$. Here it is difficult to make calculation in general case. Therefore, for example, we consider the case of $h/J=1$, $s=1$, $N=4$ and obtain that
$v_{min}=\vert J\vert\gamma\sqrt{19/2}$ if $\theta'=3\pi/4$, $\phi'=\phi$, and $v_{max}=\vert J\vert\gamma\sqrt{67/2}$ if $\theta'=\pi/4$, $\phi'=\phi-\pi$.

\section{Conclusions \label{sec8}}

We studied the evolution of $N$ arbitrary spin-$s$ system described by long-range $zz$-type Ising model having started from the initial state projected
on the positive direction of the unit vector. The final state depends on two spherical angles which define the initial state and on the period
of time of evolution. We calculated the Fubini-Study metric with respect to these parameters (\ref{form10}). This fact allowed us to analyse the
geometry of manifold which contains all states reaching during the evolution. We obtained that the parameters of metric tensor do not depend on the
azimuthal angle. This means that the manifolds with different azimuthal angles have a some geometry. So, we explored the
geometry of a two-parametric manifold with a predefined azimuthal angle. From the analysis of the Riemannian curvature and periodic conditions on the state parameters
we conclude that the manifold is closed and has the topology of the sphere. Also for $N>2$ and $s=1/2$ or for $s>1/2$ the manifold has a
dumbbell-shape structure. It has the concave part with the centerline with respect to polar angle.

The speed of evolution of the system on quantum state manifold was studied. It was obtained the conditions on the initial state for achieving the maximal
and minimal values of speed. The dependence of the scalar curvature on the speed of evolution was obtained (\ref{form20}). This fact allows to experimentally
measure the curvature of quantum state manifold through the speed of evolution. We suggested the physical system of methane molecule for application of
these considerations. The four protons of hydrogen atoms have a tetrahedral structure that allows of their spins mutually interact.
So, for this system we obtained the behavior of the speed of evolution and the dependence of scalar curvature on this speed.

Also we studied the geometry of the quantum state manifold and speed of evolution of the system in the thermodynamic limit ($N\to\infty$).
For this purpose we divided the coupling constant $J$ by the number of spins, so that the energy per spin remains finite. Such rescaling
does not influence the form of scalar curvature of the quantum state manifold. Therefore, in this case the scalar curvature is also determined by
equation (\ref{curvature2}). It is easy to see that in the thermodynamic limit the scalar curvature tends to zero in all points of the state manifold
except the points with $\theta=\pi/2$. So, an increase in the number of spins of the system leads to the inflation of the state manifold and
it becomes flattened. Also, due to the dumbbell-shape structure of state manifold, it has the concave part, which in the thermodynamic limit becomes
the closed line at $\theta=\pi/2$. The scalar curvature of all points on this line takes the value $-16/\gamma^2$. Such structure of the quantum
state manifold causes the divergence of the speed of evolution in all points of the manifold except the points with $\theta=\pi/2$.
At these points, in the thermodynamic limit the speed of evolution diverges as $\sqrt{N}$. On the line with $\theta=\pi/2$ the speed of evolution
has a finite value $\vert J\vert\gamma s/\sqrt{2}$.

Finally, we explored the influence of an external magnetic field directed along the some axis on the metric of state manifold. It was obtained
that the presence of magnetic field changes the metric of manifold. However, we do not know anything about the topology of this
manifold (except the case when magnetic field is directed along the $z$-axis) because the evolution of the system becomes very complicated.
In this case we can argue that under the influence of the magnetic field there are no points on manifold where $g_{\chi\chi}=0$. So, this means
that the eigenstates of the system do not belong to the manifold. Also we analysed the influence of magnetic field on the topology of manifold
when the magnetic field is directed along the $z$-axis. In this case we obtained that when the ratio between the values of magnetic field and interaction
coupling is irrational number then the manifold has a topology of infinitely long cylinder, otherwise the manifold remains similar to the sphere.

Also the conditions on magnetic field (\ref{form27}) which allow to reach the minimal possible speed of evolution (\ref{form26}) for predefined initial
state was found. As an example, for some predefined initial states the conditions for achieving the minimal and maximal speed of evolution were calculated.

\section{Acknowledgements}
The author thanks Prof. Andrij Rovenchak for useful comments.
This work was supported by Project FF-30F (No. 0116U001539) from the Ministry of Education and Science of Ukraine.

\newpage

\begin{appendices}
\section{Derivation of the metric tensor\label{appa}}
\setcounter{equation}{0}
\renewcommand{\theequation}{A\arabic{equation}}

In this appendix we in detail explain how to calculate the metric tensor of the manifold defined by the state (\ref{form6}). This state depends
on three real parameters $\chi$, $\theta$ and $\phi$ which determine the manifold of quantum states. Each point on this manifold corresponds
to the state with predefined set of parameters. To obtain the components of metric tensor first of all
we calculate the derivatives of state (\ref{form6}) with respect to parameters $\chi$, $\theta$ and $\phi$ and then combine the scalar products
which are included in the definition (\ref{form8}). So, the derivatives have the form
\begin{eqnarray}
&&\vert\psi_{\chi}\rangle =-i2\sum_{1\leq i < j\leq N}S_i^zS_j^z e^{-i2\chi\sum_{1\leq i < j\leq N}S_i^zS_j^z}e^{-i\phi\sum_{i=1}^NS_i^z}e^{-i\theta\sum_{i=1}^NS_i^y}\vert s,s,\ldots,s\rangle\nonumber\\
&&\vert\psi_{\theta}\rangle = -ie^{-i2\chi\sum_{1\leq i < j\leq N}S_i^zS_j^z}e^{-i\phi\sum_{i=1}^NS_i^z}\sum_{i=1}^NS_i^ye^{-i\theta\sum_{i=1}^NS_i^y}\vert s,s,\ldots,s\rangle\nonumber\\
&&\vert\psi_{\phi}\rangle = -i\sum_{i=1}^NS_i^ze^{-i2\chi\sum_{1\leq i < j\leq N}S_i^zS_j^z}e^{-i\phi\sum_{i=1}^NS_i^z}e^{-i\theta\sum_{i=1}^NS_i^y}\vert s,s,\ldots,s\rangle.
\label{derivatives}
\end{eqnarray}
Using equations (\ref{spincomp}) and (\ref{BCHeq}) we easily calculate the scalar products
\begin{eqnarray}
&&\langle\psi\vert\psi_{\chi}\rangle =-iN(N-1)s^2\cos^2\theta,\quad \langle\psi\vert\psi_{\theta}\rangle=0,\quad \langle\psi\vert\psi_{\phi}\rangle=-iNs\cos\theta,\nonumber\\
&&\langle\psi_{\chi}\vert\psi_{\chi}\rangle=N^2s^4(N-1)^2\cos^4\theta+2N(N-1)^2s^3\sin^2\theta\cos^2\theta\nonumber\\
&&+\frac{1}{2}N(N-1)s^2\sin^4\theta,\quad \langle\psi_{\chi}\vert\psi_{\theta}\rangle=-iN(N-1)s^2\sin\theta\cos\theta,\nonumber\\
&&\langle\psi_{\chi}\vert\psi_{\phi}\rangle=N^2(N-1)s^3\cos^3\theta+N(N-1)s^2\sin^2\theta\cos\theta,\nonumber\\
&&\langle\psi_{\theta}\vert\psi_{\theta}\rangle=\frac{Ns}{2},\quad \langle\psi_{\theta}\vert\psi_{\phi}\rangle =\frac{i}{2}Ns\sin\theta,\nonumber\\
&&\langle\psi_{\phi}\vert\psi_{\phi}\rangle=N^2s^2\cos^2\theta+\frac{1}{2}Ns\sin^2\theta.
\label{scalarprod}
\end{eqnarray}
Now substituting these products in definition (\ref{form8}) we obtain the metric tensor (\ref{form10}).

\section{Calculation of the angular defect \label{appb}}

In this appendix we obtain the formula (\ref{form12}) that defines the contribution of the angular defects in the Gauss-Bonnet theorem (\ref{gaussbonnet}).
The manifold defined by metric tensor (\ref{form10}) has a two singular points at $\theta=0$ and $\pi$. Near this points the manifold has an angular
defects which have the form of cones. It is easy to see if we write the metric for these areas. Since these areas are close to the singular points,
we can write the metric to the second order of $\theta$ (\ref{form11}). So, using this considerations, let us obtain the equation which allows to calculate
the angular defect near the singular point. The angular defect can be defined as a difference between the angle $2\pi$ and the angle which the system
passed during the one period around the singular point
\begin{eqnarray}
\Delta=2\pi-\frac{s_{\chi_{max}}}{r},
\label{angledefectcon}
\end{eqnarray}
where $s_{\chi_{max}}$ is the distance which the system passes during this period
and $r$ is the distance between the system trajectory and the singular point. To calculate $s_{\chi_{max}}/r$ we use the formula (\ref{pass}) with $g_{\chi\chi}$ from metric (\ref{form11}).
Thus, during the one period ($t=\chi_{max}/J$) the system passes the following distance
\begin{eqnarray}
s_{\chi_{max}}=\sqrt{g_{\chi\chi}}\chi_{max}=\sqrt{2}\gamma (N-1)s\sqrt{Ns}\theta\chi_{max}.
\label{distancearsp}
\end{eqnarray}
Using $g_{\theta\theta}$ from metric (\ref{form11}) we can write $r$ in the form
\begin{eqnarray}
r=\sqrt{g_{\theta\theta}}\theta=\gamma\sqrt{\frac{Ns}{2}}\theta.
\label{conegeneratrix}
\end{eqnarray}
So, substituting expressions (\ref{distancearsp}), (\ref{conegeneratrix}) in definition (\ref{angledefectcon}) and taking into account that the manifold
defined by metric (\ref{form10}) has a two angular defects we obtain formula (\ref{form12}).

\end{appendices}

\end{document}